\documentclass{jfm}

\usepackage{epsfig}
\usepackage{graphicx}
\usepackage{natbib}
\usepackage{color}
\usepackage{soul}

\ifCUPmtlplainloaded \else
  \checkfont{eurm10}
  \iffontfound
    \IfFileExists{upmath.sty}
      {\typeout{^^JFound AMS Euler Roman fonts on the system,
                   using the 'upmath' package.^^J}%
       \usepackage{upmath}}
      {\typeout{^^JFound AMS Euler Roman fonts on the system, but you
                   dont seem to have the}%
       \typeout{'upmath' package installed. JFM.cls can take advantage
                 of these fonts,^^Jif you use 'upmath' package.^^J}%
      }
  \else
  \fi
\fi

\ifCUPmtlplainloaded \else
  \checkfont{msam10}
  \iffontfound
    \IfFileExists{amssymb.sty}
      {\typeout{^^JFound AMS Symbol fonts on the system, using the
                'amssymb' package.^^J}%
       \usepackage{amssymb}%
         \let\leq=\leqslant
         
      }{}
  \fi
\fi

\ifCUPmtlplainloaded \else
  \IfFileExists{amsbsy.sty}
    {\typeout{^^JFound the 'amsbsy' package on the system, using it.^^J}%
     \usepackage{amsbsy}}
    {}
\fi

\newsavebox{\astrutbox}
\sbox{\astrutbox}{\rule[-5pt]{0pt}{20pt}}

\title{High-speed X-ray imaging of a ball impacting on loose sand}

\author{Tess Homan$^1$, Rob Mudde$^2$, Detlef Lohse$^1$, and Devaraj van der Meer$^1$}
 
\affiliation{
$^{1}$Physics of Fluids, 
University of Twente, The Netherlands\\
$^{2}$Department of Chemical Engineering, Delft University of Technology, The Netherlands\\}

\date{17 September 2014} 

\begin{document}

\maketitle

\begin{abstract}
When a ball is dropped in fine, very loose sand, a splash and subsequently a jet are observed above the bed, followed by a granular eruption. 
To directly and quantitatively determine what happens inside the sand bed, high-speed X-ray tomography measurements are carried out in a 
custom-made setup that allows for imaging of a large sand bed at atmospheric pressures. 
Herewith we 
show that  the jet originates from the pinch-off point created by the collapse of the air cavity formed behind the penetrating ball.
Subsequently we 
measure how the entrapped air bubble rises through the sand and show that this is  
consistent with bubbles rising in continuously fluidized beds.
Finally, we measure the packing fraction variation 
throughout the bed. 
From this we show that there is (i) 
a compressed area of sand in front of and next to the ball while the ball is moving down, 
(ii) a strongly compacted region at the pinch-off height after the cavity collapse; and (iii) a relatively loosely packed center in the wake of the rising bubble.
\end{abstract}

\maketitle

\section{Introduction}
Impact events --violent collisions of objects with targets-- are ubiquitous in nature and industry. They range from 
the impact of asteroids on planet surfaces \citep{Melosh1989}, raindrops falling onto soil or sea \citep{Worthington1908}, to the interaction of metal droplets and melts in the metallurgical industry. Air entrapped during impact events may be beneficial --allowing for the aeration of the sea-- or detrimental, when gas bubbles are trapped in liquid steel. In both cases it is crucial to understand the processes that take place below the surface of the target.

When an object impacts on a deep layer of water a splash is formed and a few milliseconds later a jet shoots out of the water. 
Because of the transparent nature of water it is possible to directly observe what happens below the surface~\citep{Worthington1908,Oguz1990,Lohse2003,Bergmann2009}: 
While the intruder moves through the water layer, an air cavity is formed. 
The walls of the cavity move toward each other due to hydrostatic pressure. 
At the moment the cavity walls collide two jets are formed, one going up and one going down. 
The air-pocket entrapped 
below the pinch-off point moves down with the intruder, detaches, and then slowly rises to the surface. 

For impacts onto non-transparent media such as sand direct imaging is not feasible. Upon impact of a steel ball on a bed of fine, very loose sand 
a splash and jet appear 
above the surface~\citep{Thoroddsen2001,Walsh2003,Lohse2004PRL,Lohse2004Nature,Ciamarra2004,Royer2005,Boudet2006,Royer2007,Caballero2007,Katsuragi2007,Marston2008,Seguin2008,Umbanhowar2010,Pacheco2011,Ruiz_Suarez2013} that are very similar to those observed for water. 
The questions that arise are: What happens below the surface of the granular bed and to what extent is this similar to the sequence of events in water?

To answer these questions we must ``look'' inside the sand bed, for which we use a unique, custom-built high-speed X-ray tomographic setup. Previously, high speed X-ray imaging has been used by Royer~\emph{et al.} in a series of pioneering papers~\citep{Royer2005,Royer2011}, using a very different setup that makes use of a parallel X-ray beam. Due to restrictions on the X-ray apparatus used, those experiments were conducted in a 
setup that is smaller than the setups used in~\citep{Lohse2004PRL,Lohse2004Nature,Caballero2007,Royer2008} which can lead to unwelcome boundary effects~\citep{vonKann2010}. For the same reason, and similarly with possible side-effects, the silica particles (sand) of~\citep{Lohse2004PRL,Lohse2004Nature,Caballero2007,Royer2008} were substituted by Boron Carbide particles.  


In this paper we 
present 
impact experiments done in a custom-made high-speed X-ray tomography setup which is large enough to allow for the direct study of the experiments described in~\citep{Lohse2004PRL,Lohse2004Nature,Royer2005,Caballero2007,Royer2008}, \emph{i.e.}, in the original size and using the same silica sand bed. 
In fact, we will 
show that  the jet originates from the pinch-off point created by the collapse of the air cavity formed behind the penetrating ball. In addition, 
we 
measure the density changes in front of the sphere and in the pinch-off region during the collapse of the bed. Finally, we observe how the  entrapped air bubble rises through the sand bed. 
In the next section the experimental setup will be introduced, whereafter we describe three different ways to analyze the data.
First the air-cavity and jet formation will be reconstructed in `Cavity reconstruction'.
In the `Rising air bubble' section we will take a close look into the shape and rising mechanism of the air bubble.
Last, in `Local packing fraction', the density changes of the sand around the ball and the air-cavity are explored.

\section{Experimental setup}
A cylindrical container that is 1~m high and with an inner diameter of 15~cm is filled with sand until a certain height $H$ (see Fig.~\ref{xray:fig:setup}). 
The bottom of the container consists of a porous material to allow for fluidization of the sand and the container is fully closed. 
An electromagnet is suspended from a rod, such that a metal ball (diameter $d=3$~cm) can be released from different heights. 
Before every experiment the sand is fluidized to destroy the existing network of contact forces, and subsequently the airflow is turned off slowly to allow the sand to settle into a very loose state. 
The height of the sand bed above the plate after fluidization ($H$) and the release height ($h$) are measured. 
The size distribution of the individual sand grains (with density $\rho=2.21 \pm 0.04$~g/cm$^3$) is between 20~$\mu$m and 40~$\mu$m and the average packing fraction after fluidization is 0.41.
For this research several sets of experiments were carried out with 
different experimental conditions. The key parameter we varied is the impact velocity $v_i$ and can be described 
using the Froude number $\textrm{Fr}\equiv 2 v_{i}^2/gd$, or equivalently, $\textrm{Fr}=4(h-H)/d$. Varying the Froude number from 9 to 92 resulted in the same qualitative behavior. 
Therefore, in this paper, we do not extensively discuss the influence of this parameter, but merely state which experimental condition is used.

The container is placed in an unique custom-built X-ray setup~\citep{Mudde2008,Mudde2010Powder,Mudde2010Industrial} (also shown in Fig.~\ref{xray:fig:setup}), which consist of three powerful X-ray sources (\emph{Yxlon}, 133-150keV, 4 Amp.) with three arrays of detectors placed in a triangular configuration. 
A single detector bank consists of two horizontal rows (spaced 40~mm apart) of 32 detectors that are positioned on an arc such that the distance between the source and the detectors is constant at 1386~mm. 
Each detector consists of an CdWO$_4$ scintillation crystal ($10\times 10\times 10$~mm) coupled to a photo diode and the data is collected with a sampling frequency of 2500 Hz. 
Two sets of experiments were carried out: 
One in which the container was positioned in the center of the X-ray setup and measurements are taken by all three detector banks, and a second set of experiments where the container was placed close to one of the sources (at a distance of 275~mm) for enhanced spatial resolution. In this last set measurements are taken from only one detector bank.

Each detector measures the attenuation of the X-rays on the path between the source and the detector. 
The attenuation of single wavelength X-rays is described by the Lambert-Beer law~\citep{Mudde2010Powder,Mudde2010Industrial}, which states that there is a logarithmic dependence between the number of registered photons per second and the absorption coefficient of the specific material times the path length. 
In this problem the only parameters that change are the path length through sand and the path length through air, where the latter can be neglected in this setup due to the very low absorption of X-rays by air. 

Every single detector is calibrated such that it gives the length of prepared sand on the path between the source and the detector, $l_s$. 
As a first point in the calibration we used a fully fluidized bed. 
Note that because the container has a circular cross-section the length of the path through the sand varies for different detectors.
Next, we place a rectangular container filled with air inside the bed in the path of the rays, and again prepare the bed.
This changes the amount of sand in-between the X-ray source and the detectors.
The boxes are made of very thin plastic to minimize their influence on the X-ray signal.
When the exact position of the containers is known it is possible to calculate the equivalent path length (\emph{i.e.}, the length of the path the X-ray travels through the sand, as calculated from the measured signal) for each detector. 

\section{Cavity reconstruction}
\label{sec:cavrec}
How can we reconstruct what happens inside the sand when a ball impacts? 
With the setup described above we measure the response of the bed in one horizontal cross-section as a function of time. 
Because we are interested in the complete cavity shape within the bed, the experiment will have to be repeated while measuring at different heights, $z$. 
The results can later be stitched together. This method requires that the experiment is very reproducible. 
To check this we first examine the center detector signals for several repetitions of the experiment at an average depth within the sand while the ball penetrates the bed.

\subsection{Center detector}
\begin{figure}
\begin{center}
\includegraphics[width=86mm]{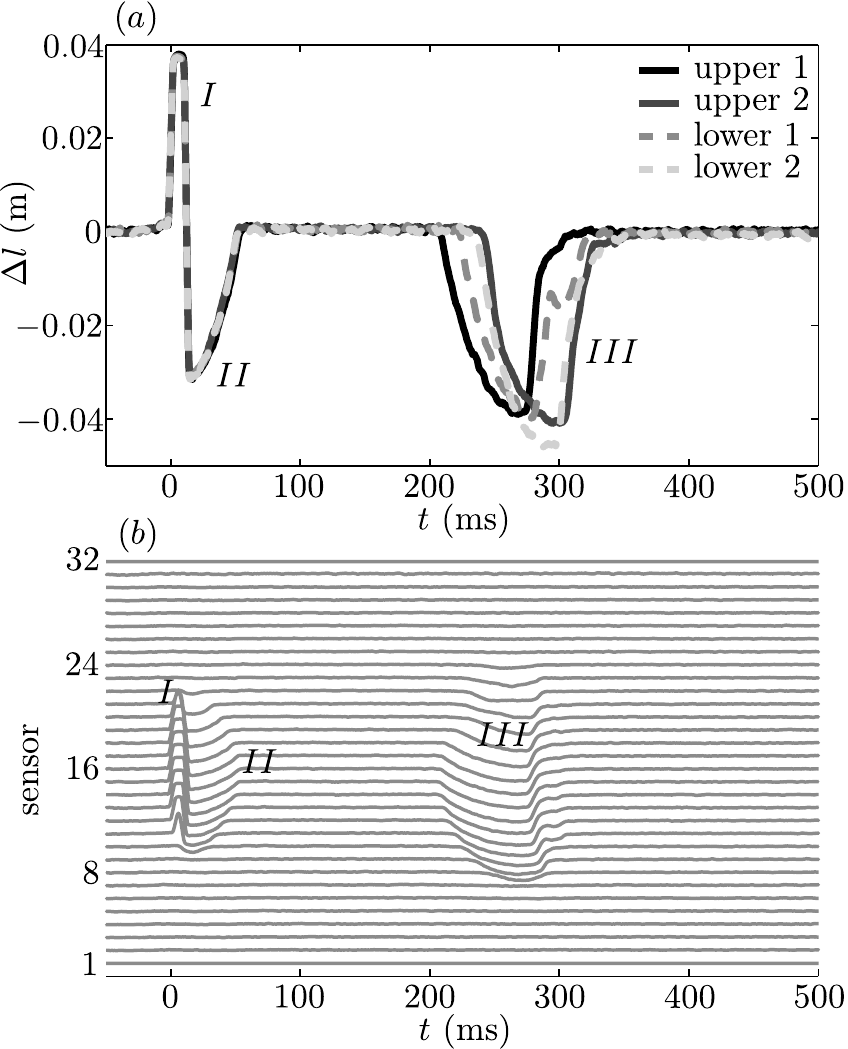}
\caption{\label{xray:fig:reproducibility} $(a)$~The measured signal of the center detector as a function of time for four different realizations of an impact experiment using the same measurement height ($z$=8~cm below the surface, $\textrm{Fr}=92$). 
Two of the measurements are recorded with the upper detector bank and two measurements with the lower detector bank. 
When the ball passes through a ray the signal becomes higher, whereas an air cavity accounts for a lower signal. 
The first part of the signal, which corresponds to the passing of the ball ($I$), cavity creation, and cavity collapse ($II$), is very reproducible. 
The second part of the signal, corresponding to the rising air bubble ($III$), shows poor reproducibility.
$(b)$~The measured signals of all the sensors of one detector array measured in a single experiment at one fixed height, $z$. The number of detectors that see the ball~(I) or the air cavities~(II and III) provides an estimate of the size of the object. Similar-sized objects that are visible in the signal for longer time move with a lower velocity through the measurement plane.
}
\end{center}
\end{figure}

The signals of the center detector for different experiments at a fixed height are shown in Fig.~\ref{xray:fig:reproducibility}a. 
On the vertical axis the change $\Delta l$ of the equivalent path length (\emph{i.e.}, compared to the situation before impact) is plotted during the impact. 
When the ball passes through the measurement plane the signal of the central detectors drops due to the higher absorption coefficient for X-rays of metal compared to sand. 
This leads to an increase of $\Delta l$ ($I$). 
Immediately after the ball passes, the signal $\Delta l$  becomes negative ($II$), indicating that in the path of the ray there is less sand than would fit in the container, \emph{e.g.}, as would happen when an air cavity has formed in the wake of the ball.
This implies that there is more air in the sand at this height, but it does not reveal how this air is distributed. 
The bed may have become very loose such that the air is evenly distributed, or the air may be concentrated in the center as an air cavity. 
Some time later a negative peak again suggests the presence of air ($III$).

\begin{figure}
\begin{center}
\includegraphics[width=86mm]{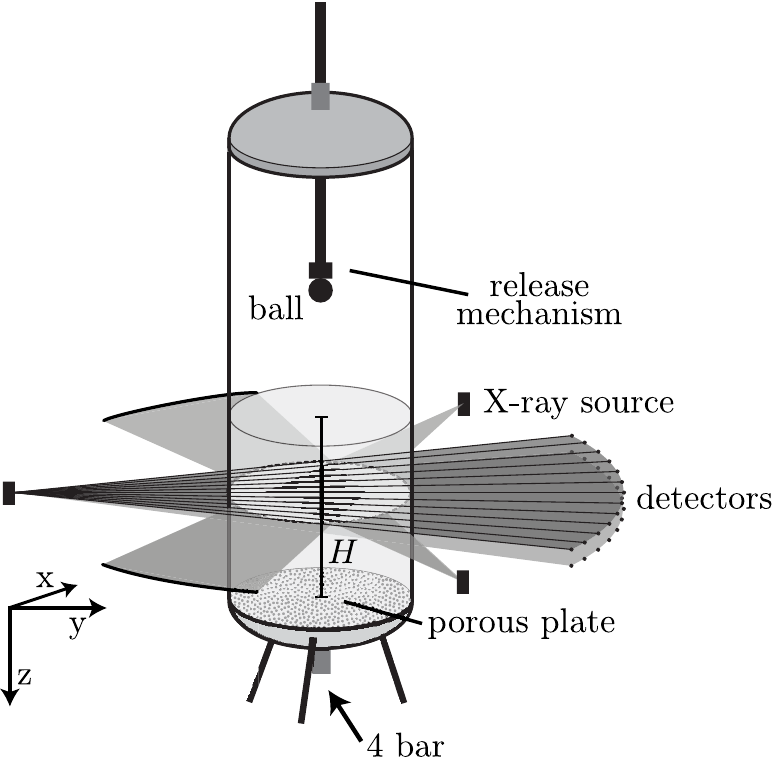}
\caption{\label{xray:fig:setup} Schematic of the setup used in the X-ray experiments. The setup consists of a container filled with very fine sand. Near the bottom a porous plate is mounted such that air can be blown in, fluidizing the sand. A ball is dropped from various heights using an electromagnet into a loosely settled bed. The setup is placed in a custom-made tomographical X-ray device consisting of 3 X-ray sources and 6 arrays of 32 detectors: Opposite to each X-ray source two arrays of detectors are placed in one detector bank (only one bank is shown).}
\end{center}
\end{figure}

The different curves show four distinct realizations of the experiments all recorded at the same measurement plane in the sand. 
Two of them are measured with the upper detector row and two of them with the lower array of detectors. 
The first part is very reproducible, which can be concluded from the fact that the equivalent path lengths and the duration of the peaks are equal. 
The second part, 200~ms to 400~ms after impact, is less reproducible. 
The measured values are similar, but the shape and timing of the peak are very different among different experiments. 
From all of the above we can deduce that it must be possible to accurately reconstruct the impact within the sand bed, up to a certain amount of time after the ball has impacted ($t\leq200$~ms).
This timespan must at least be sufficient to image the formation of the jet, judging from the time scale ($t<100$~ms) on which the latter forms.

\subsection{Cavity size and shape}
Since data is available from an entire array of detectors it is possible to obtain information about sizes and positions. 
The signals from the different detectors of one of the arrays are plotted above each other in Fig.~\ref{xray:fig:reproducibility}b. 
The number of sensors that detect the ball (the positive part of the signal) reflect the width of the ball. 
The negative signals are concentrated in the center of the container indicating that the additional air exists in the form of an air cavity rather than somehow dispersed through the sand bed. 
These negative signals are found immediately after the ball passes ($II$) showing that the air cavity is attached to the ball. 
Since a similar number of detectors ``see'' both the cavity and the ball, the air cavity must have a width similar to that of the ball. 
The second air cavity ($III$), the signals of which arrive at the detectors after a considerable delay, is visible in more sensors than the ball, demonstrating a larger size of this second cavity, which can be interpreted as a detached air bubble rising through the sand bed.
Note that an object moving at a lower velocity will have a longer X-ray signal duration because it is longer in field of view of the detector.
The fact that the signal duration of the air bubble is longer than that of the air cavity does not necessarily imply that the air bubble is bigger; the magnitude of the signal however does give information about the size.

\begin{figure}
\begin{center}
\includegraphics[width=\textwidth]{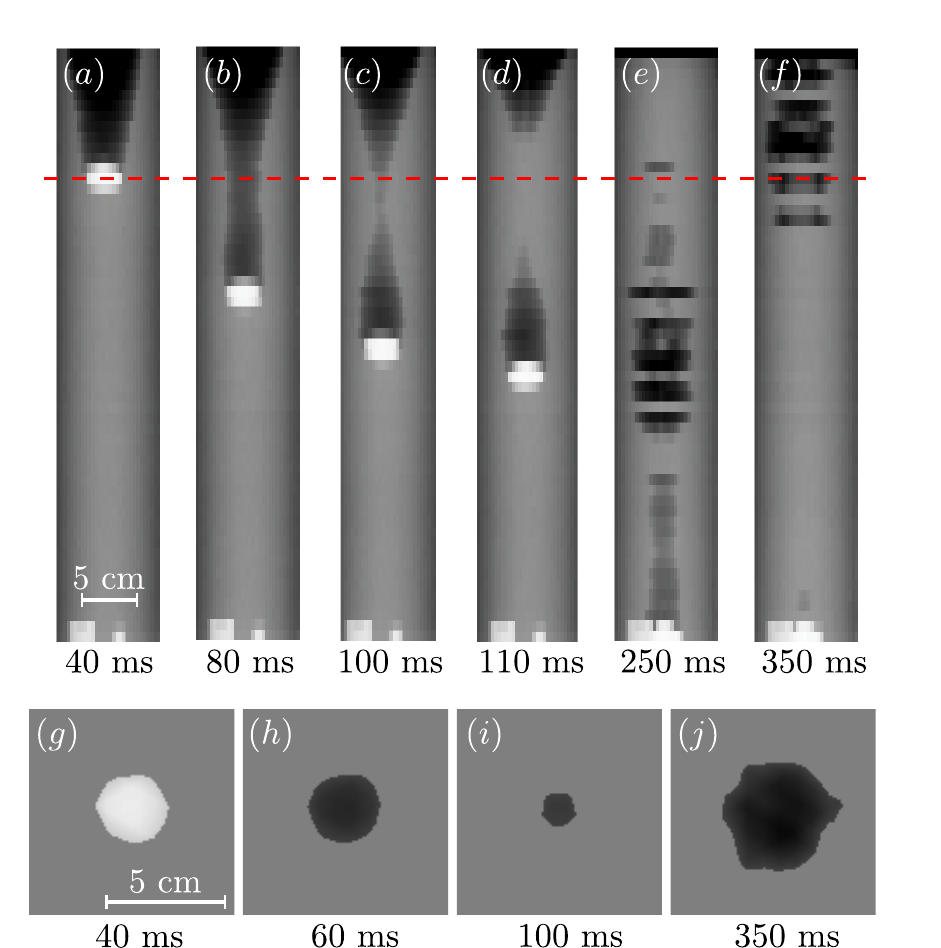}
\caption{\label{xray:fig:density} $(a-f)$ A series of plots at different times after the ball has impacted onto the sand bed at $t=0$ for $\textrm{Fr}=92$. The grey-scale value represents the normalized signal. For each plot, the horizontal axis displays the signals of the different detectors and the vertical axis repetitions of the experiment with cross-sections taken at different depths in the bed. We clearly see the cavity first being formed and subsequently closing~$(a,b)$, the resulting pinch-off~$(c)$, the formation of the jet, and finally the entrapment of an air bubble in the sand~$(d)$. In the next plot~$(e)$ the air bubble detaches from the ball and slowly rises to the surface~$(f)$.
$(g-j)$ Tomographic reconstruction of a single horizontal cross-section through the bed, see red dashed line in top figure, at 4 different times for $\textrm{Fr}=89$.
From left to right: a measurement plane through the center of the ball~$(g)$, through the air cavity immediately behind the ball~$(h)$, the air cavity close to the collapse~$(i)$, and a cross-section through the rising air bubble~$(j)$.
}
\end{center}
\end{figure}

In Fig.~\ref{xray:fig:density}a-f the change in equivalent path length is plotted as a grey scale value (white for positive and black for negative $\Delta l$) for different times during the experiment. 
The pixels in each row indicate the signal of the different detectors at a single height, whereas the different rows correspond to experiments done at different depths in the bed. 
This gives a first indication of what happens inside the sand bed. 
As the ball moves through the sand an air cavity behind the ball is generated~$(a)$. 
This air cavity grows while the ball moves and then starts to collapse under the influence of the lithostatic pressure (i.e., the ``hydrostatic pressure'' due to the mass of grains above that point) in the bed~$(b,c)$.
When the walls of the cavity touch~$(d)$, a jet shoots upward and an air bubble is entrained. 
The air bubble moves down with the ball and after it detaches it slowly rises to the surface~$(e,f)$. 
From this analysis it is clear that the events in~$(a-d)$ are highly reproducible, whereas the randomness and irregularity in the last two plots~$(e,f)$ reflects that the rising of the air bubble is not reproducible at all.
The size of the latter varies considerably with height, indicating that in the different experiments to which they correspond the bubble detaches from the ball at different points in time, leaving empty gaps in the reconstruction.
The critical question that remains is: To what extent are the cavities and resulting jets that are created axisymmetric? 
To obtain more insight into this issue, we will now look at the full tomographic information that is available from the setup.

\subsection{Tomography}
When the container is positioned in the center of the X-ray setup, a single horizontal cross-section can be imaged by three detector arrays spaced evenly around the contianer, allowing for tomographic measurements. 
Using \mbox{tomography}, we obtain the full 2D shape of the cross-section of the ball and the air-cavities.
In Figs.~\ref{xray:fig:density}g-j the tomographic reconstruction at a single height, for four different times during the experiment is shown.
For the tomography the signals of all three detector banks are super-positioned on a square lattice of 140 by 140 pixels.
By applying a threshold intensity value the cavity and the ball shape can be extracted.
In the first image~$(g)$ the ball is visible, of which the size and shape are known.
Indeed, within the limits of the reconstruction the ball is found to be round, and also the size is correctly estimated. 
In Fig.~\ref{xray:fig:density}h a reconstruction of the air cavity just after the ball passed is provided. 
The air cavity is seen to have a similar degree of roundness as the ball, and has the same size as the ball, which is indeed what one would expect directly after the cavity is created.
The second image of the air cavity~$(i)$ is taken just before the collapse, showing that the air cavity is still axisymmetric. 
The last reconstruction~$(j)$ is made at the time the bubble passes by. 
The bubble is not completely circular, and is rising slightly off-center.
This is one of the origins of the poor reproducibility of the air bubble.

\subsection{A single array}
\begin{figure}
\begin{center}
\includegraphics[width=86mm]{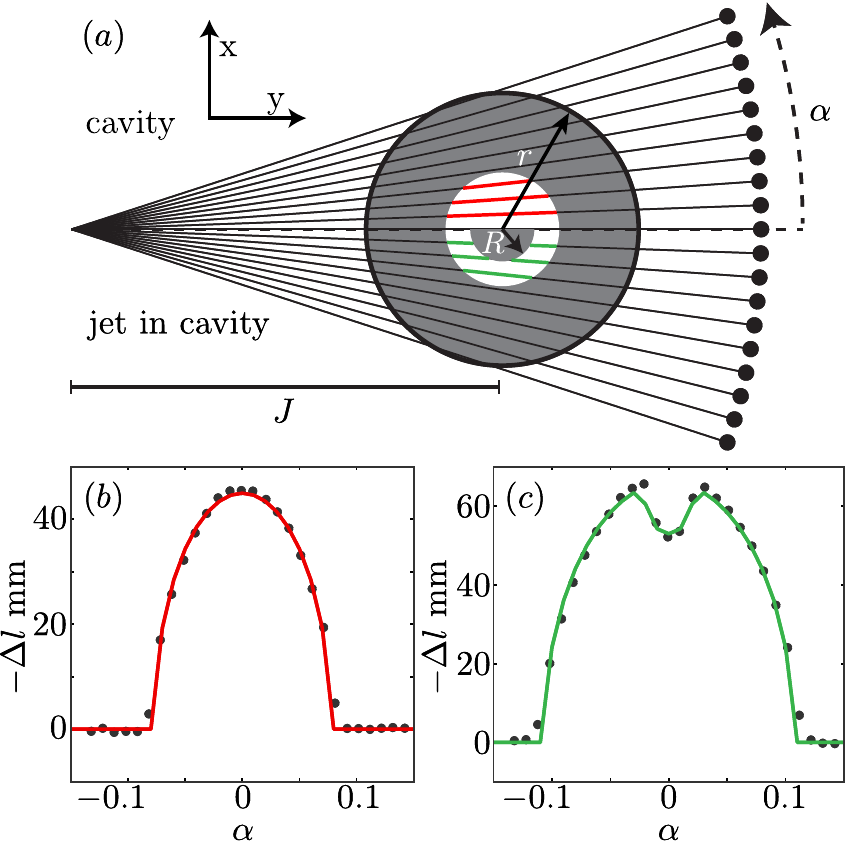}
\caption{\label{xray:fig:fit} $(a)$~Geometry of one horizontal cross-section of the bed. When the exact locations of the source and detectors with respect to the container are known it is possible to calculate the change of the equivalent path length $\Delta l$ as a function of the angle $\alpha$ for both the case with a circular air cavity (upper half) and the situation in which a sand jet is present in the center of the air cavity (lower half). In $(b,c)$~the equivalent path length $\Delta l$ is plotted as a function of the angle $\alpha$ for two situations: $(b)$~an air cavity in the sand bed and $(c)$~a jet within the air cavity. Both signals are fitted to the theoretical case of a circular cavity and a circular jet.}
\end{center}
\end{figure}

By moving the setup closer to one of the X-ray sources we are able to obtain a much higher spatial resolution.
From the tomography analysis there are good indications that the cavity remains axisymmetric at least until the collapse.
For such an axisymmetric cavity we can anticipate what the signals in the different detectors should look like, given the radius of the cross-section of the cavity. This is illustrated in the top half of Fig.~\ref{xray:fig:fit}a. 
To quantify the cavity shape as a function of time the cavity radius for every measurement (each height) needs to be determined at each point in time. 

In Fig.~\ref{xray:fig:fit}b, for a given time and height, the measured equivalent path length is plotted as a function of the angle between the detector and the center detector when there is an air cavity present. 
This data can be fitted with the expectation for an axisymmetric air cavity in the center of the container, as shown in the top half of Fig.~\ref{xray:fig:fit}a. 
The equivalent path length of a circular cavity of radius $r$ as a function of the angle $\alpha$ is calculated to be

\begin{equation}
\Delta l(r,\alpha)=-2\sqrt{\frac{(r^2-J^2)\tan{\alpha}^2+r^2}{\tan{\alpha}^2+1}},
\label{xray:eq:ltheoretisch}
\end{equation}
where $J$ is the known distance between the X-ray source and the center of the container.
This function is fitted to the obtained data to get the cavity radius $r$, red line in Fig.~\ref{xray:fig:fit}b.

After the collapse a jet occurs inside the air cavity in some of the cross-sections, as illustrated in the bottom half of Fig.~\ref{xray:fig:fit}a.
This will change the signal as shown in Fig.~\ref{xray:fig:fit}c, where we observe a shape similar to that of Fig.~\ref{xray:fig:fit}b, but with a pronounced dimple in the center.
To calculate both the cavity radius and the jet radius, equation~(\ref{xray:eq:ltheoretisch}) is adapted such that the change in equivalent path length is calculated through two concentric circles. 
The larger circle is filled with air and the smaller one filled with sand: $\Delta l(r,\alpha)-\Delta(R,\alpha)$, where $R$ is the radius of the jet. 
A fit for both $r$ and $R$ is plotted as a green line in Fig.~\ref{xray:fig:fit}c.

\begin{figure}
\begin{center}
\includegraphics[width=\textwidth]{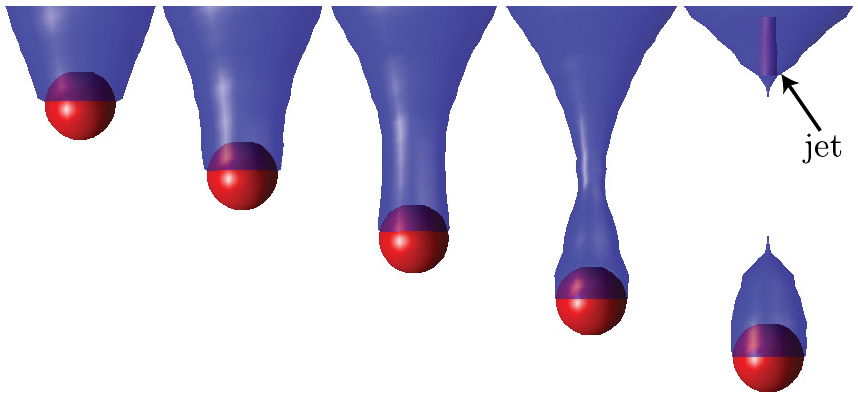}
\caption{\label{xray:fig:3D} Several snapshots of the results from the analysis described in Section~\ref{sec:cavrec}. To obtain these images the equivalent path lengths measured in the X-ray setup are fitted to theoretical cavity shapes. Stitching together the experiments executed at different heights results in the blue air cavity. The ball position is also measured from the data and the ball is added to the images in red. For the last image the jet recreated from the data is visible in purple. The plots represent the situation 40, 60, 80, 100, and 120~ms after the impact for $\textrm{Fr}=92$}.
\end{center}
\end{figure}

This fitting procedure is repeated for every time step and every measurement height, such that we are able to reconstruct the full axisymmetric cavity- and jet-shape as a function of time. 
The result of this analysis is shown in Fig.~\ref{xray:fig:3D}.
In blue the cavity radius as a function of height is represented in the figure at several times after impact. 
The exact position of the ball could be extracted from the original data by looking at the maximum (see Fig.~\ref{xray:fig:reproducibility}).
The ball is plotted in Fig.~\ref{xray:fig:3D} in red.
Despite its small scale we were also able to reconstruct the jet that is created during the collapse.
The last plot of Fig.~\ref{xray:fig:3D} shows the jet in purple.

From the analogy to cavity collapse in a liquid, a secondary, downward moving jet is expected to form from the pinch-off point as well \citep{Lohse2004PRL,Bergmann2009}. We find no clear evidence for this secondary jet, which may be connected to the fact that it is expected to be weaker than the upward one such that it is simply too thin to measure with the spatial accuracy of our experimental setup.

\subsection{Cavity collapse}
With the analysis described above the cavity radius is extracted as a function of time. It is now possible to take a closer look at the dynamics of the cavity collapse, both at and below the closure depth.

Collapsing air bubbles in incompressible liquids have been studied both theoretically and experimentally \\citep{Eggers2007,Gordillo2006,Gekle2009,Duclaux2007,Gekle2009,Bergmann2006}. 
Theory has shown that the time evolution of the cavity radius asymptotically and slowly converges to a power law with exponent $1/2$. More specifically, the local slope $\alpha=d\ln r/d\ln t$ has been shown to satisfy $\alpha \approx 1/2 + 1/[4\sqrt{-\ln (t_c-t)}]$~\citep{Eggers2007,Gordillo2006,Gekle2009,Duclaux2007}. 
In experiments and numerics this leads to data that over several decades appear to be extremely well-fitted by a power law~\citep{Gekle2009,Bergmann2006}, with a measured exponent that is slightly larger than $1/2$.

\begin{figure}
\begin{center}
\includegraphics[width=0.8\textwidth]{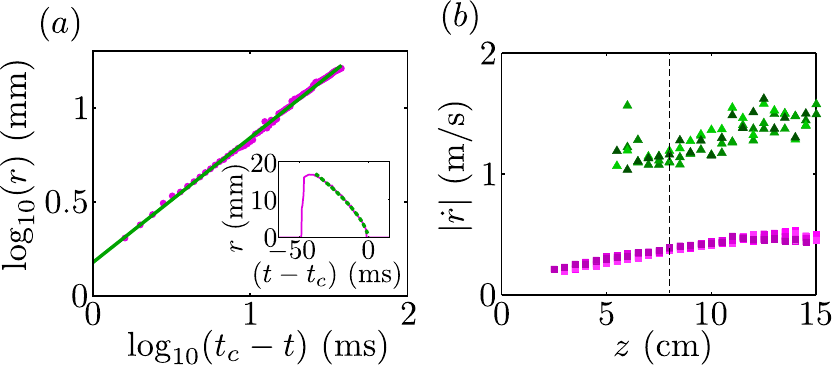}
\caption{\label{xray:fig:collapse} $(a)$~Double logarithmic plot of the cavity radius as a function of time at $z=8.0$~cm below the surface, which is the closure depth for $\textrm{Fr}=92$. The cavity is created immediately after the ball has passed and in the beginning has roughly the same radius as the ball. As the cavity is collapsing, the signal decreases. The inset shows the same data in a linear plot. $(b)$~Closing velocity $|\dot{r}|$ as a function of depth $z$, $(i)$ estimated from dividing the distance and the duration of the collapse (starting from the maximum cavity radius, squares) and $(ii)$ calculated from a power-law fit as in $(a)$, evaluated at a radial distance of $r=20$~mm from the symmetry axis (triangles). The different shades of color indicate four different experimental realizations at this Froude number and the vertical dashed line marks the pinch-off point.}
\end{center}
\end{figure}

Is the behavior similar for a granular pinch-off, and what is the underlying mechanism? 
In Fig.~\ref{xray:fig:collapse}a the cavity radius is plotted as a function of time on a double logarithmic scale, with the time given relative to the closing of the void at $t_c$, \emph{i.e.}, 
$t_{c}-t$. 

The behavior is clearly consistent with a power law during the collapse of the cavity.
The best fit exponent (green line) gives $r \sim (t_c-t)^{0.66}$, but due to limited resolution close to pinch-off we are not able to quantitatively compare with values found by Gekle~\emph{et al.}~\citep{Gekle2009} for the void collapse in water. 
However, values for the exponent as high as $0.66$ have also been found in the liquid impact experiment~\citep{Bergmann2006} which suggests that the mechanism of cavity collapse in a granular bed is quite similar to that in liquids. More specifically, our findings are consistent with the model of radial cavity collapse initiated by a hydrostatic (or lithostatic) driving pressure and continued by the inertia of the medium, as was already suggested in~\citep{Lohse2004PRL}.

The final question we address in this Section is: How does the cavity collapse away from the point of first closure. To answer this question, in Fig.~\ref{xray:fig:collapse}b we plot the absolute value of the radial closing velocity $|\dot{r}|$ as a function of the depth $z$, estimated in two ways. The first method we use is a global one in which we divide the maximum cavity radius by the time interval from the time this maximum is reached to cavity closure (squares in Fig.~\ref{xray:fig:collapse}b). The second one is local and makes use of power-law fits of the type $r(t)=a(t_c-t)^b$ for every depth $z$ from which the  velocity $\dot{r}$ at a fixed distance $r_0=20$~mm is calculated as $\dot{r}=a(b-1)(t_c-t)^{b-1}=a(b-1)[r_0/a]^{(b-1)/b}$ (triangles). Both methods are consistent in the sense that they give the same trend with depth, but as expected the second methods provides larger values for $|\dot{r}|$ since the velocity diverges towards $r=0$. 
In any case, clearly, the results of both methods show that $|\dot{r}|$ increases slowly with depth. 

In the literature there has been some discussion about secondary, lower pinch-offs that may be responsible for the thickening of the lower part of the jet \citep{Royer2005,Royer2008,vonKann2010}. In the context of this discussion our present result indicates
that a second --deeper-- pinch-off is capable of creating a jet that could be almost as strong and fast as the first pinch-off. This supports the view put forward in~\citep{vonKann2010} that the thick-thin structure first reported in~\citep{Royer2005} is caused by a secondary jet catching up with the first. By bursting through the primary pinch-off region it is then assumed to create the thick part of the visible jet.\\\\     

Concluding this Section, for the Froude numbers studied in our setup ($\textrm{Fr}=9-92$) we have unambiguously shown that the jet at atmospheric pressures originates from the primary pinch-off point of the cavity, and not from the pressurized air bubble as was suggested in~\citep{Royer2005} based upon Xray experiments in a much smaller setup. In fact the entrained air bubble is observed to move down with the sphere (see Figs.~\ref{xray:fig:density}d and \ref{xray:fig:3D}) until it detaches and starts rising through the sand bed. This latter process will be discussed in the next Section. 

The experimental technique used in this paper opens the possibility of directly observing the formation of the thick-thin structure reported in refs.~\citep{Royer2005,vonKann2010} at reduced ambient pressure. However, from high-speed imaging experiments in the current setup it turned out not to be possible to observe these structures at atmospheric pressures, such that additional research at reduced ambient pressures is needed for this purpose.

\section{Rising air bubble}
A remaining question regarding the impact events is the mechanism by which the detached air bubble moves towards the surface.
An intuitive way of thinking about rising air bubbles in a granular medium is that the unsupported grains on top of the bubble ``rain'' down through the center into a pile at the bottom.
This transport of material will give the bubble a net upward velocity.
A second mechanism, often used for continuously fluidized beds, is closer to the rising of air bubbles in water.
Material from the perimeter of the bubble is transported along the interface towards the bottom of the bubble, where a wake is formed.

In our experiment we do measure a rising bubble, but due to poor reproducibility of this part of the experiment we cannot stitch the different experiments at different heights together to obtain a spatial image (see Figs.~\ref{xray:fig:reproducibility} and~\ref{xray:fig:density}).
To reconstruct the bubble shape we therefore have to find a different method to analyze the data.
Because the bubble is moving in the vertical direction, in due time the entire bubble will pass any horizontal cross section.
This means that if the velocity of the bubble is known it is also possible to retrieve the shape of the bubble from a single experiment done at one height, such that we don't have to worry about reproducibility.

As shown in Fig.~\ref{xray:fig:setup} we record the data with two detector arrays simultaneously.
In a single experiment the air bubble will pass the two measurement planes that go from the X-ray source to the upper and lower array of detectors.
The distance between these two planes in the center of the container is 4~mm.
By determining the time difference of the front and back of the bubble passing the two measurement planes we obtain the speed of the front and back of the bubble. 
The difference in velocity between the front and the back is found to be small enough such that we can assume that the bubble rises with a constant speed.
Comparing measurements at different heights, we find no clear trend of the bubble velocity as a function of height, which is at least partially due to the poor reproducibility of the experiment in this regime.
We find that all bubble rise velocities are around $0.3\pm 0.1$~m/s.

\begin{figure}
\begin{center}
\includegraphics[width=86mm]{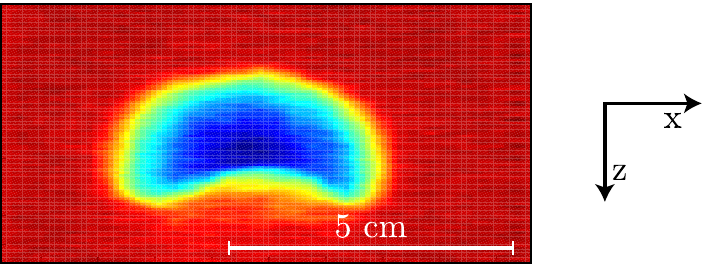}
\caption{\label{xray:fig:bubble} The shape of the rising air bubble for an experiment at $\textrm{Fr}=92$. This shape is obtained by recording the radius of the air bubble that passes by in time at a single height. Plotting the signal from the different detectors gives the complete shape. The color indicates the width of the bubble perpendicular to the paper, dark blue is a width of 4.5~cm.}
\end{center}
\end{figure}

When the time axes are rescaled with the constant bubble speed we get a bubble shape as shown in Fig.~\ref{xray:fig:bubble}.
In the horizontal direction the information from the different detectors is displayed.
The colors represent the depth of the bubble perpendicular to the plane of view.
The bubble is spherical cap shaped, like a bubble rising in a fluid, or in a continuously fluidized bed.
The bottom of the bubble is concave, which is consistent with either a pile or a wake.

In 1963 Davidson and Harrison~\citep{Davidson1963} presented a relation for the rising velocity of a single bubble in a fluidized bed: $u_b=0.71\sqrt{gd_{\textrm{eq}}}$, where $d_{\textrm{eq}}$ is the equivalent bubble diameter $d_{\textrm{eq}}=(\frac{6}{\pi}V_b)^{\frac{1}{3}}$ with $V_b$ the bubble volume.
Now that we have the shape of the bubble we can estimate the velocity using this model which gives a value of $0.44$~m/s.
This is close to our experimental value, which is slightly lower.
This stands to reason, since our bubble is not rising in a continuously fluidized bed and thus a lower velocity is expected.

We are not able to see if there is a rain of particles within the bubble, since we measure the average signal over a line instead of locally, and the density of the ``rain" would be very low.
However, the shape of the bubble and the rising velocity are close to bubbles rising in a continuously fluidized bed, suggesting that the rise mechanism will be similar as well.

The shape of our measured bubble is similar to the air bubble measured by Royer~\emph{et al.} in~\citep{Royer2005} although they have a different explanation for the shape.
They attribute the concave bottom of the bubble to an impinging second jet, that grows to meet the first jet.
We however find that the rise velocity is consistent with a rising air bubble, that will finally erupt at the surface, rather than overtaking the primary jet.

Note that the shape of the rising air bubble is very different from the shape it had when it was still attached to and dragged along with the moving ball, as can be appreciated by comparing Fig.~\ref{xray:fig:bubble} to, e.g., Fig.~\ref{xray:fig:density}(d). Finally, in all of the over $500$ repetitions of the experiment that have been measured for this work, we have never observed more than a single rising bubble.

\section{Local packing fraction}
Until now we have assumed that the packing fraction of the bed does not change significantly. 
This assumption was necessary to calculate the air path lengths in the bed. 
By simply observing the experiment it is obvious that the packing fraction must change, since when we compare the bed height before and after impact we find that it has lowered~\citep{Caballero2007}. 
From the initial very loose state that is created by the fluidization procedure we end up with a more compactified bed. 
We want to determine the corresponding change in the packing fraction, and we want to discover how the compactification is distributed throughout the bed.

The experiment provides the equivalent path length of the X-rays through the sand, $l_s$. 
Whenever a given X-ray does not encounter an air cavity in its path the change in this path length ($\Delta l$) before and after the experiment can, in first order, be related to a change in packing fraction by:
\begin{equation}
\frac{\Delta \phi}{\phi_{\textrm{before}}}=\frac{\Delta l}{l_{s,\textrm{before}}}
\label{xray:eq:packingfraction}
\end{equation}
where $\Delta \phi$ needs to be interpreted as the average packing fraction changes along the path.
Note that the values for $\Delta \phi$ given in this section correspond to much smaller $\Delta l$ than those discussed in the previous section. In fact, we measure packing fraction changes of only a few percent. This however needs to be contrasted with the maximum range that can be expected during compaction, i.e., the difference between loose and dense packings, which is typically of the order of 10 \% or less: For the experimental setup and the sand used in our experiments we have determined in a tapping experiment that the difference between the densest and the loosest obtainable packing is of the order of 10 \%. 

\begin{figure}
\begin{center}
\includegraphics[width=86mm]{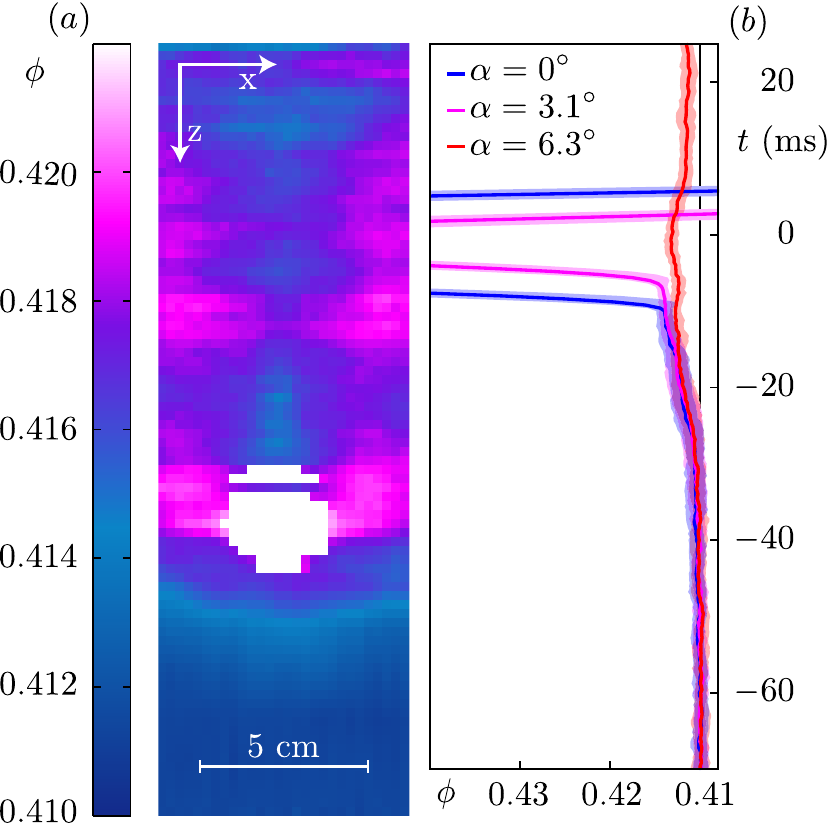}
\caption{\label{xray:fig:density_differences} $(a)$~A representation of the packing fraction of the bed after an experiment at $\textrm{Fr}=52$. The data is taken several seconds after the impact events have terminated, assuring that there are no air pockets left in the sand. The color indicates the packing fraction ($\phi$). Even though the change in packing fraction is small, a compactified area next to the ball can be observed, while the vertical strip above the ball is relatively loose. The compaction below the ball decreases with depth. $(b)$~The averaged packing fraction plotted as a function of time. To obtain the curves the signal of 20 different experiments is averaged. The three different curves give the signal from three different detectors, \emph{i.e.}, for three different values of $\alpha$. The transparent area around the curves indicates the statistical error. All three signals show a clear increase in packing fraction just before the ball the ball blocks out most X-rays.}
\end{center}
\end{figure}

\subsection{Packing fraction after the experiment}
The local packing fraction after the experiment in the entire sand bed, calculated with equation~(\ref{xray:eq:packingfraction}), is shown in Fig.~\ref{xray:fig:density_differences}a.
These measurements were taken several seconds after the ball has come to a halt, which assures that there are no air-cavities or bubbles left, and that only packing fraction variations are detected.
Note that the packing fraction before the experiment was equal to 0.41, uniformly throughout the container.
This means that the bed is compactified during the experiment.

We see a clear compacted region (pink area) next to where the ball has stopped.
The packing fraction above the ball (in the center of the plot) is relatively low, and is the lowest just on top of the ball.
The packing fraction below the ball slowly decreases with depth back to a value of 0.41.
Wherefrom do these packing fraction variations originate, and what do they teach us about the events below the surface?

\begin{figure}
\begin{center}
\includegraphics[width=\textwidth]{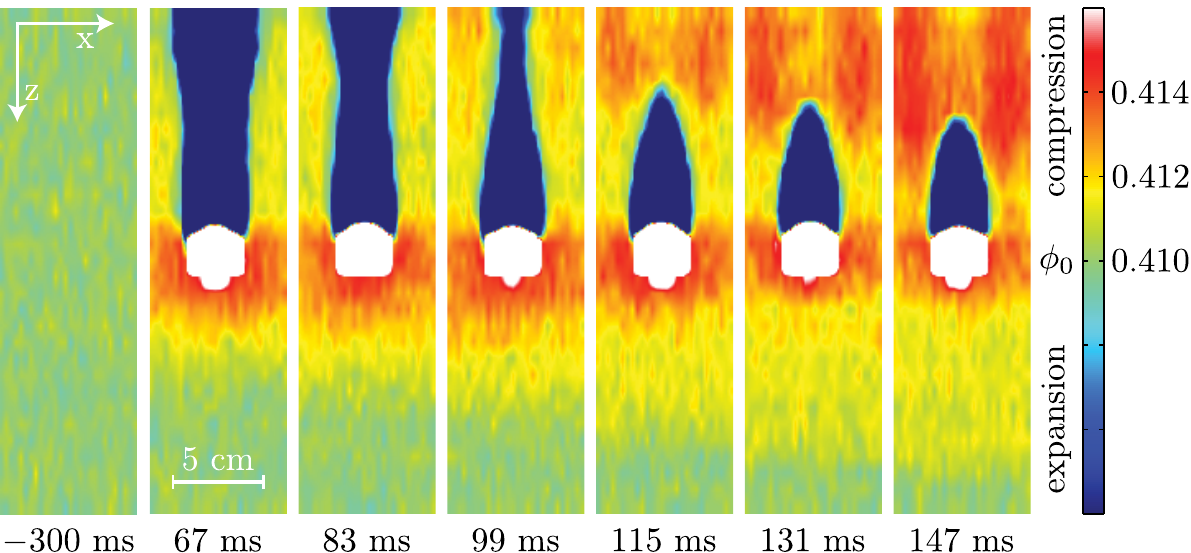}
\caption{\label{xray:fig:around_the_ball} The packing fraction around the ball at several times during impact with a Froude number of $52$. To obtain these images we move along with the ball and averaged the signals around it. The first image shows the bed before impact (average packing fraction of 0.41). In the other frames the ball moves from 8.5 cm (67 ms after impact) below the surface to 16 cm (147 ms after impact) below the surface. During this movement we observe a compacted region to the sides and in front of the ball (red), a growing compactified region (yellow) and a strong compression above the ball where the air cavity pinches off. Note that the white area does not exactly represent the measured ball shape but simply the area in which no reliable data for density changes in the sand could be calculated due to the presence of the ball.}
\end{center}
\end{figure}

\subsection{Time resolved packing fraction}
To understand the packing fraction of the sand around the ball after the experiment we need to look into the local compaction while the ball is moving through the sand.
To determine what happens with the sand just in front of the ball we zoom in on the signal before  the ball passes by at a given height. 
This gives the bed density underneath the moving ball. 
To obtain sufficient data the signals of 20 different experiments at 10 different heights are averaged. 
The moment the ball passes by is used to synchronize the signals in time. 
To smoothen the signal a central scheme is used where the trend of the 10 previous points is extrapolated beyond the central point. 
In Fig.~\ref{xray:fig:density_differences}b the result of this analysis is shown. 
The blue curve ($\alpha=0^\circ$) passes through the center of the ball and therefore detects the ball first. 
The other two curves show the signal through the side of the ball ($\alpha=3.1^\circ$) and completely beside the ball ($\alpha=6.3^\circ$). 
All three curves show a clear increase of the signal before the front of the ball passes. 
This shows that there is a compaction of the sand just before the ball arrives.
Or, there is a compacted region being pushed in front of the ball. 
From the red curve ($\alpha=6.3^\circ$) we deduce that there is also compaction next to the ball. This compacted region is still present after the experiment is finished, as can be seen in Fig.~\ref{xray:fig:density_differences}a.

\subsection{Packing fraction during penetration}
The density differences of the sand during the experiment are very small, but it is possible to detect them if the signal is averaged over a sufficiently large time-window (see Fig.~\ref{xray:fig:density_differences}b). 
To image the packing fraction variations during the penetration around the ball we switch to the frame of reference of the ball such that we are able to do a time averaging. 
The result of this procedure is shown in Fig.~\ref{xray:fig:around_the_ball}. 
The red area below and next to the ball indicates a compacted region, just as we saw in the previous section.
In time we see that the compacted region below the ball (yellow area) grows downwards relative to the ball and obtains a size that is several times that of the sphere.

When comparing these findings to the data obtained by Royer~\emph{et al.} in refs.~\citep{Royer2007,Royer2011}, who reported a much smaller compacted region in front of the ball at atmospheric pressures for comparable Froude numbers, we may attribute the difference to the higher sensitivity of our setup to packing fraction changes. 

In addition to what happens below the ball, we can also investigate the compaction above the ball.
First the air cavity is visible (blue) and when the cavity collapses a growing red area indicates a compacted region next to the pinch-off.
The data in Fig.~\ref{xray:fig:density_differences}a (taken after the experiment was done) shows a relatively uncompacted area in the center above the ball.
This must be connected to the rising bubble rearranging the sand particles in its path.
It suggests that the sand at the bottom of the bubble is deposited loosely, pointing to a slow and unpressurized mechanism.

\section{Conclusion}
Using a custom-made high-speed X-ray tomography setup we measured the events that occur below the surface when a ball impacts on a bed of fine, very loose sand.
We were able to reconstruct the air cavity until and beyond the collapse by stitching together measurements done at different depths. 
From the cavity reconstruction we learned that 
the phenomena below the surface are similar to the events that occur during and after an impact in water: A cavity is formed behind the penetrating ball,
the cavity collapses, creating a jet and entraining an air-bubble. Even the power-law behavior with which the cavity collapses 
is consistent with the pinch-off of an air cavity in a liquid. Using the signal of a single experiment done at one height we were 
able to retrieve the shape of the rising air bubble. Both the shape and the rising velocity of the bubble are very similar to those of bubbles rising in a continuously fluidized bed.

During the experiment the sand bed is compactified. Even though the change in the signal caused by the compaction is very small compared to that from the air cavities, we were able to measure sand that is compressed in front and to the side of the ball while the ball moves through the bed.  
This compacted area grows in time and typically has a size of several ball diameters when the ball comes to rest. This is much larger than the compacted region reported at atmospheric pressures in refs.~\citep{Royer2007,Royer2011}, which may have implications for explaining the pressure dependence of the drag entirely from the ``cushioning'' effect put forward in these papers. 
The compaction does decrease with increasing distance from the ball, and is most pronounced in the center of of the container, below the ball. Moreover, during the cavity collapse the sand at the collapse height is also greatly compressed and in the last step (the rising of the air bubble) the sand has time to rearrange itself:  With the deposition of sand at the bottom of the bubble we end up with a relatively loose center and compacted sides. 
Just above the position where the ball stops we have the area with the lowest compaction.
This is where the air bubble pinches off from the ball, giving rise to the existence of a region that is depleted of sand grains.

\bibliographystyle{jfm}

\end{document}